# Machine Agency in Human-Machine Networks; Impacts and Trust Implications

Vegard Engen, J. Brian Pickering and Paul Walland

IT Innovation Centre
Gamma House, Enterprise Road
Southampton, SO16 7NS, UK
{ve,jbp,pww}@it-innovation.soton.ac.uk

**Abstract.** We live in an emerging hyper-connected era in which people are in contact and interacting with an increasing number of other people and devices. Increasingly, modern IT systems form networks of humans and machines that interact with one another. As machines take a more active role in such networks, they exert an increasing level of influence on other participants. We review the existing literature on agency and propose a definition of agency that is practical for describing the capabilities and impact human and machine actors may have in a human-machine network. On this basis, we discuss and demonstrate the impact and trust implications for machine actors in human-machine networks for emergency decision support, healthcare and future smart homes. We maintain that machine agency not only facilitates human to machine trust, but also interpersonal trust; and that trust must develop to be able to seize the full potential of future technology.

## 1 Introduction

The term social network is widely understood given the proliferation of communication platforms such as Facebook and LinkedIn. However, non-human component(s) are often neglected. In some cases, this may simply be because they are not obviously visible to end-users. However, increasingly sophisticated ICT systems form networks of both human and machine participants. Machines are not just passive participants in such networks, merely mediating communication between humans; they are increasingly adopting an active role, enabled by technological advances that allow greater autonomy and the performance of increasingly complex tasks.

Here, we refer to a Human-Machine Network (HMN) as a collective structure where humans and machines interact to produce synergistic and often unique effects. Humans and machines are both viewed as actors, given that they interact with at least one other actor in the system. Referring to machines/objects as actors/agents raises issues with agency, which is an ongoing debate in the literature as reviewed in Section 2.

While most psychology and sociological models only attribute agency to human actors, more recent models have been proposed that attribute agency also to machines, such as Actor-Network Theory (ANT) [1] and the Double Dance of Agency (DDA) model [2]. Nevertheless, agency definitions still seem insufficient, as Jia *et al.*. [3] argue in the context of the Internet of Things (IoT). Moreover, the scope of research on machine agency is limited. The DDA model was proposed against a background of Information Systems (IS) research and based only on empirical work on Enterprise Resource Planning (ERP) use cases, while Jia *et al*. [3] focus only on the IoT. Other models, such as captology [4] have a wider scope in terms of empirical applications, but focus solely on human-to-computer interaction.

In this paper, we draw upon the existing literature and propose definitions of human and machine agency (in Section 3) for the purpose of understanding and designing HMNs. Further, we discuss the relationship between machine agency and trust. Although traditionally a person-to-person construct, trust in technology has been shown to be accommodating towards technology failings [5, 6]. Further, as a form a social capital, trust may be regarded as an overall organising principle [7].On this basis, human actors do demonstrate an ability to adapt their concept of trust in response to machine agency [3, 4]. We explore three case studies in Section 4 to give a broader view of agency and its impact on the interactions between the different actors.

This work is part of the HUMANE project (http://www.human2020.eu), which is creating a typology and method for the design of HMNs. The aim of this work is to improve public and private services by uncovering and describing how new configurations of HMNs change patterns of interaction, behaviour, trust and sociability. Agency forms part of the HUMANE typology [8], which aims at improving system designers' understanding of the potentially complex interactions that may take place, particularly in terms of emergent phenomena and unexpected behaviour, ultimately leading to HMN designs that contribute innovation and creativity to future network collaborations.

## 2 Background

In respective sections below, we provide further background on what we mean by an HMN, and review the existing research on agency.

### 2.1 Human-Machine Networks

We refer to an HMN as a collective structure where humans and machines interact to produce synergistic and often unique effects. Humans and machines are both viewed as actors, given that they interact with at least one other actor in the system. Existing research relevant to HMNs has been conducted in three different areas: Socio-Technical Systems (STS) [9], Actor-Network Theory (ANT) [1] and Social Machines (SM) [10].

As discussed in [11], each of the three areas focus on particular aspects of HMNs: STS on business systems design, providing a body of theory that covers complex interactions between people and technology; ANT on social systems as networks comprising a range of heterogeneous elements, both human and non-human, viewing the actors as networks themselves; SM on online systems mediating social interactions between human users. In this paper, we focus on human-machine interaction, rather than machine mediated human-human interaction.

## 2.2 Agency

Early cognitive models of human behaviour were based on mechanical and computational processes. These were subsequently modified to include a socio-cultural element [12], whereby contextual interactions could influence responses in the short and long term. This view has evolved into theoretical models seeing human beings as conscious and creative, operating according to belief systems that enable them to set and achieve goals that drive their behaviour and avoid undesirable situations in their lives.

There are different theoretical models and frameworks that address this topic, offering different definitions of agency, which occasionally conflict. For example, in structuration theory (ST), agency is defined as "the capability to make a difference" [13]; in social cognitive theory (SCT) "agency refers to acts done intentionally" [12]. Bandura [12] describes agency according to four attributes: intentionality, forethought, self-reactiveness and self-reflectiveness. Intentionality encompasses the ability to choose to behave in a certain way; in particular with a future course of action in mind. Forethought is related to the temporal characteristic of intentionality, in the sense of setting goals, which affects behaviour in order to achieve desired outcomes and avoid undesired ones. Moreover, it is linked to motivation, which guides the chosen actions and anticipations of future events, which "provides direction, coherence, and meaning to one's life" [12].

We focus on intentionality as it is a key differentiator for definitions of agency. Compared to humans, machines do not have self-generated intention or motivation, do not experience trust or reliance, and do not behave altruistically or irrationally of their own volition; they have no agency in that sense [3, 14, 15]. However, there is a need to refer to machine agency, as machines can participate in HMNs in increasingly significant roles, influencing other actors and the outcomes of the HMN itself [2, 15, 16].

Theoretical positions have been proposed that attribute agency also to machines, such as ANT [1] and the double dance of agency model [2]. As noted above, ANT is a theory that gives equal weight to machine actors compared with human actors [1]. Moreover, Law [1] argue that sociological theories are lacking in fully understanding social effects due to excluding non-human actors, both machines and architectures. Further, ANT only refers to the agency of the network, not of independent actors.

Rose et al. [2, 15, 16] have studied machine agency in the context of Information Systems (IS), finding that neither the definitions in ST or ANT are appropriate for practical application. Rose et al. [16] argue that agency should be acknowledged for both humans and machines, like in ANT. However, unlike ANT, they recognise that their agency is different, due to properties such as self-awareness, social awareness and intentionality, which are considered exclusive to human agency [16].

Rose and Truex [15] propose in their earlier work that machines have perceived agency; neither accepting or denying the notion of machine agency. This relates to the human tendency towards anthropomorphism [17]. Rose et al. [16] also assert agency to machines in terms of the intention of their creators via their endowed potential actions. More research on this aspect forms part of captology, which is the study of persuasive technologies; defined as "interactive technolog[ies] that changes a person's attitudes or behaviours" [4]. Fogg [4] refers to intentionality as a requirement for persuasion, though acknowledging that machines do not have intentions. This semantic predicament is addressed in captology by defining a technology as persuasive when they have been created for the purpose of changing people's attitudes or behaviours [4].

Rose et al. [2] propose a 'Double Dance of Agency' model based on observing the intertwined nature of human and machine agency. That is, there are emergent outcomes stemming from the process of humans and machines interacting and that human agency itself responds to and shapes machine agency [2]. Moreover, Jia et al. [3] argue that both human and machine agency can be balanced and enhanced in order to achieve improvements in interactions and user experience.

In summary, the existing literature on agency has evolved significantly over time. Earlier definitions of agency are bound to the social context in which machine agency does not exist. Later, researchers have argued that machine agency does exist, but that it is different to human agency, and probably the result of human interaction and perception. Building on the existing literature, in the following section we propose an updated definition suitable for the analysis and design of HMNs.

## 3  Agency in Human-Machine Networks

In the context of designing, evaluating or studying HMNs, we argue the need to consider both human and machine actors in terms of agency. We aim at a) identifying a shared understanding of both human and machine agency in HMNs, while b) delineating differences between the two.

Broadly speaking, we understand the agency of an actor, whether human or machine, as the capacity to perform activities in a particular environment in line with a set of goals/objectives that influence and shape the extent and nature of their participation. The environment in this context is bound by the HMN.

SCT distinguishes between three modes of agency [12]. The first, direct personal agency, covers the four characteristics focused on in Section 2.2. The second mode, proxy agency, refers to socially mediated agency in which an agent utilises desired resources or expertise in other agents to act on their behalf to achieve goals they cannot achieve on their own. This alone may be the key motivation for human users to participate in a particular HMN, and they may exercise proxy agency via both human and machine agents. Some time after SCT was originally proposed, it was extended to the third mode; collective agency [18]. This mode is based on the premise that many goals are only achievable via socially interdependent effort. Moreover, Bandura [12] notes that group attainments are largely due to interactive, collaborative and synergistic dynamics of the interactions between the people exercising collective agency. We emphasise the importance of synergistic dynamics here as it is integral to the definition and value of an HMN (see Section 2.1).

In the context of a HMN, we can refer to actors' having different degrees of agency. For example, the agency of human actors is typically constrained by the respective ICT system of which they form a part and the interfaces they can use for interacting with the system and other actors, e.g., to exercise proxy agency. In practical terms, we can scope agency of both human and machine actors in HMNs according to three key factors: (a) the activities the actor can perform, (b) the nature of the activities, and (c) the ability to interact with other actors.

**The activities the actor can perform (a)** typically differ between the actors in an HMN; perhaps varied according to roles and responsibilities in the HMN. For example, some may be restricted to viewing content while others may be allowed to generate content, and moderators in online communities typically have higher levels of agency than other community members. In many HMNs, individual machine actors may have low agency as they may be focused on performing a limited number of fixed tasks, e.g., peripheral devices such as sensors.

**The nature of the activities (b)** concerns actors' ability to behave diversely, unpredictably, freely, and creatively in order to pursue their respective goals/objectives for participating in the HMN. We can distinguish between open or closed activities. A closed activity is one that is restricted or fixed; hence, predictable with an expected predefined outcome. An open activity is one in which the actors are able to exercise a degree of freedom, leading to diversity in the HMN. For example, allowing actors to express themselves in free text allows creativity and unpredictability in their activities. Although actors are limited to communicate via 140 characters, Twitter enables individuals and organisations to socialise, perform marketing, customer relationship management, engage in political propaganda, etc. Moreover, machine actors are effectively able to do the same activities as can human actors. While some machine actors are under direct control by human actors to, e.g., schedule tweets to propagate news, other machine actors (bots) act and pose as humans [19, 20] in order to exert influence, e.g., by spreading misinformation and propaganda [19].

**The ability to interact with others (c)** determines an actor's potential for influencing other actors or the working of HMN itself. The ability to influence others depends on the degree to which the actors are able to communicate with each other and how open the form of communication is. There are examples of HMNs where users may not even be aware that they are in fact a part of a network, such as reCAPTCHA [21]. Nevertheless, they form part of a network of millions of users, mediated via a machine, though they are not visible to one another, and, thus, do not exert influence.

Human agency is important to consider when analysing or designing a HMN as it reflects the ability of the human actors to be creative and use the HMN in unforeseen ways. It also links to the motivations of users. If they are not able to exercise agency in the HMN, in terms of achieving the objectives that underpin their reasons for participating, they may choose not to participate.

While machines cannot exhibit true direct personal agency, due to factors such as intentionality, as discussed above, they can exhibit agency in different ways. For example, it is useful to refer to machine agency in terms of the intentions of their human designers, as interactive technologies may be deployed to change human attitudes or behaviours [4]. For example, in the field of affective computing, emotionally intelligent technologies are developed to respond and adapt to users emotional needs [22].

In practical terms, our definition of machine agency reflects the degree to which machine actors may a) perform activities of a personal and creative nature (e.g., supporting health care by personalising motivation strategies), b) influence other actors in the HMN, c) enable human actors to exercise proxy agency, and d) the extent to which they are perceived as having agency by human actors. Higher levels of machine agency imply a need to consider the implications of the machine's role in the HMN, which relates to, e.g., the trust relationship between humans and machines. This is discussed further in the following section.

## 4 Machine Agency and Trust

Any relationship between human and machine agents must be based on trust and reliance in response to trustworthiness factors in machines [23, 24]. Humans may not simply apply social metaphors to machine services [25], although trust transfer, once developed in one area, often occurs across others [7, 26]. Whilst Corritore et al. [27] stress both affective and cognitive dimensions of technology trust, Wiegmann et al. [28] maintain that trust between human agents becomes reliance on machines. Either way, human agents show greater tolerance and adaptation, even in the face of failing or inadequate technology [5, 6, 29]. The interplay between trust and trustworthiness in complex HMNs, especially in light of increasing machine agency and emergent behaviours, therefore merits specific investigation.

### 4.1 An Evacuation Perspective

Mistrust between operational staff and the public may cause problems [30, 31], whereby operational and emergency services misinterpret crowd behaviours with negative [32], even catastrophic, consequences [33]. The eVACUATE project is going some way to solve this problem (http://www.evacuate.eu). Human actors within the HMN include the members of the public (potential evacuees), operational staff responsible for the general safety of a venue, and, in extreme cases, members of emergency services (fire brigade, ambulance; special forces; etc.). The human and machine actors operate at different levels of autonomy and, therefore, agency; and relations between and within groups may vary contextually from indifference to significant dependence. Therefore, trust becomes an increasingly significant factor. Operational staff and emergency services are reliant on machines such as sensors, smart devices and decision-support engines, while in an emergency, evacuees may vitally depend on them.

The HMN must provide constant data on crowd movement as well as the immediate environment to facilitate not only decision making by those responsible, but the mediation of trust between all actors. For if the original Mayer et al.. model is correct [23, 24], then the three main antecedents of trust – ability, integrity, and benevolence – are easily exposed through the interaction between machine and human agents. Sensors, such as CCTV cameras, but also smart objects like floor-level lighting and other signage, can be assumed to provide continuous and objective information back to those responsible (ability). The dynamic and targeted intervention of the latter will be clearly appreciated for what it is: communication based on fact from the sensors (integrity) with the sole purpose of indicating safe and effective evacuation routes (benevolence). Further, if personal devices such as smart phones are recruited as part of the network they can provide alerts and updates which individuals may choose to share with those around them, promoting social cohesion during an emergency, also having the potential to receive additional and more specific contextual information from evacuees about their immediate vicinity. A task-specific exchange is possible allowing the rapid creation of mutually supportive and trust-related connection between evacuees and those tasked with their safety. The HMN in the eVACUATE case therefore has enormous implications for redressing trust issues from previous disasters [34–36] while mediating intergroup communication and cooperation.

### 4.2   Am I Just A Number?

The potential for HMNs to provide mutual support, promoting not only safety and efficiency but also reciprocal trust, are clear not least in enabling communication between actors whilst at the same time providing each with additional information and support. Within healthcare and specifically in machine-mediated patient-doctor communication, the situation is different from the evacuation case. Here, as well as the medical team and patients themselves, an HMN might include monitoring devices such as wearables, mobile applications for self-reporting, a data store, and the related software to interrogate or summarise the data sensibly. The promise is in gathering data reliably and consistently to support specialist nurses and consultants, giving them an overview for their patient during appointments, and freeing them up to focus on the social and personal condition of the patient.

As an IT as a Utility (ITaaU) project, TRIFoRM [37] focused on trust development in technology in the management of chronic conditions such as rheumatoid arthritis. Participants in a small pilot included both patients and a clinician who were asked in semi-structured interviews about their willingness to trust a mobile monitoring application which would automatically log their movements, sending collated information back to the supporting clinical team. Patients acknowledged the benefits of automated tracking: between appointments, they may not always remember exactly how they felt or how things may have changed. Further, if their condition caused specific cognitive impairment, they felt the technology would fill the gaps. Technology is trusted therefore as a reliable mechanism for data collection: its ability to collect and transmit. However, this is not the whole story.

Not least since health data are considered sensitive [38] or special-category [39], surely patients would be concerned at what happens to their data, how it is curated, and who gets to see it? When asked, however, participants showed no such concern: any integrity issues were overwhelmed by an altruistic willingness to share their own data to help others. They trust the technology components in the HMN, but they were concerned that human agents might not continue to provide the interpersonal care they sought. Machine agency within a network should support and enhance trust-related interactions between the human actors.

### 4.3 How Can You Help Me?

There is much talk about the emergence of smart homes and the potential for home management systems to take over the efficient running of the systems with which our homes are becoming filled. Current home systems are automation systems rather than human machine networks since in essence the machine component is an intermediary, rather than a contributor to the performance of the network. Of interest to us here are future smart home networks in which the machine components exhibit agency, playing an active role in decision making and monitoring, thus, in order to consider a smart home as a true HMN we need to consider the case that devices in the home can learn behaviour, can suggest activity and can operate with some level of autonomy. Such a network would incorporate sensors, controllers and actuators, but also would incorporate "smartness" at the level of interaction with the human occupants and interpretation and management of the output of the various sensors. It is this added component of "smartness" that will distinguish future home networks as HMNs in the sense in which we use it here. Recent reports [40] have highlighted the role of trust and security in home networks, and indeed have suggested that the key driver for uptake of such technology in the near future will be on the basis of security fears, but in order for those fears to be allayed, the owners of smart homes will need to have a high level of trust in the operation of the system and the security of the data which it contains.

In order to consider the key features of an HMN in the home, we need to postulate a format which has not yet arrived, in which the future smart home is one in which the human home dweller communicates directly with the home systems via an intelligent interface, establishing a dialogue with a quasi-intelligent machine agent which forms a proxy for "the home". Thus the home owner would interact with home networks and functions through an interface that operates as an assistant and concierge, tracking movements and collecting behavioural data from the occupants in order to customise its behaviour and responses to changing situations.

Note that, in reality this intelligent interface is a construct representing the home server, data store and connected sensors, actuators and signage. However, the humans in the network relate to the personification of the smart home functions, attributing agency to it. This level of anthropomorphism has been shown to increase levels of both trust and, perhaps surprisingly, tolerance [5, 6, 29].

The family members comprising a household can be regarded as a single human entity, since they are maximally interconnected individuals. They each have personal communication devices for interacting with the world outside, which are also used for communication with the house interface. In turn, this has the effect of putting the house on a similar perceptual level to the humans with which they interact. This contributes to the anthropomorphism by which the human actors are able to apply concepts of trust to the home agent despite its machine nature.

The smart home network of the future can therefore be seen as an example of a human-machine network in which the human actors, the intelligent machine actors and the responsive machine actors occupy different niches, and between which different forms of trust relationship exist. In order for such a network to function effectively we are obliged to postulate a machine agent which can be anthropomorphised and imbued with human characteristics, allowing it to be part of an established trust relationship.

## 5    Discussion

From health and safety monitoring to the smart gadgets in our homes, the increasing dependence on sophisticated technology implies a fresh look at concepts such as agency. A simple assumption that machines are no more than deterministic automata confined to well-defined tasks is no longer valid. Just as teamwork and crowdsourcing increase human potential [41, 42], the question now becomes how to exploit the power of technology and move toward the creation of synergistic capabilities through human-machine interaction in HMNs.

As a first step, something approaching an equal footing in agency terms between the machine and human actors goes some way to open up the debate. Machine agency is, as we have seen, a much contested idea, and may yet struggle to escape the confines of human perceptions of usefulness and anthropomorphism. Yet, advanced decision-support systems are beginning to demonstrate a level of agency that rapidly becomes a crucial factor in critical situations. Such agency may even lead to a concern that other human actors within a network respond by withdrawing their social concern with other human actors: this may not be machine intentionality, but it certainly demonstrates the possibility for machines to make a difference [13] for human-to-human interactions.

Accepting interactive collaboration as a real possibility, HMNs such as smart homes and advanced socio-technical robotics enable social engagement, encouraging the evolution of mutually supportive networks. For this to become a real and lasting possibility will require the development and maintenance of trust. Only on a trust basis, including a willingness to compromise, to forgive and learn how to overcome shared problems, will the full potential of HMNs become a reality. To become a reality, a revision of the original definition of agency is long overdue, not least to allow the full capabilities of sophisticated technology to combine and develop together in socially motivated HMNs limited only by human imagination.

## 6    Conclusions and Future Work

We have proposed a definition of agency and discussed its practical application to analysing and designing human-machine networks (HMNs). In support of recent literature, our definitions recognise that machines can both enhance and constrain human agency as well as exhibit agency themselves. Three case studies demonstrated the importance of considering machine agency when analysing or designing HMNs in terms of influencing motivation, participation and trust. As a construct, they suggest not only the traditional view of trust as a prerequisite for technology adoption, but also a mediating role for the machine actors themselves.

Evaluation of the definitions of human and machine agency offered here is already in progress within the HUMANE project. Based on scales used to identify the degree of agency, the challenge is to establish descriptive labels which are both intuitive and useful to system designers. Further, the consistent interpretation of these scales is needed if comparisons are to be made between HMNs with a view to sharing design patterns and experience. However, in considering potential social and behavioural impact as we have done here, it is hoped that some level of prediction about how HMNs will evolve can be attempted.

**Acknowledgements.** This work has been conducted as part of the HUMANE project, which has received funding from the European Union's Horizon 2020 research and innovation programme under grant agreement No 645043.

## References

1. Law, J.: Notes on the Theory of the Actor-Network: Ordering, Strategy, and Heterogeneity. Syst. Pract. 5, 379–393 (1992)

2. Rose, J., Jones, M.: The Double Dance of Agency: A Socio-Theoretic Account of How Machines and Humans Interact. Syst. Signs Actions. 1, 19–37 (2005)

3. Jia, H., Wu, M., Jung, E., Shapiro, A., Sundar, S.S.: Balancing human agency and object agency: an end-user interview study of the internet of things. In: Proceedings of the 2012 ACM Conference on Ubiquitous Computing. pp. 1185–1188. ACM (2012)

4. Fogg, B.: Persuasive computers: perspectives and research directions. In: The SIGCHI conference on Human Factors in Computing. pp. 225–232 (1998)

5. Lee, J.D., Moray, N.: Trust, control strategies and allocation of function in human-machine systems. Ergonomics. 35, 1243–1270 (1992)

6. Lee, J.D., See, K.A.: Trust in automation: Designing for appropriate reliance. J. Hum. Factors Ergon. Soc. 46, 50–80 (2004)

7. McEvily, B., Perrone, V., Zaheer, A.: Trust as an organizing principle. Organ. Sci. 14, 91–103 (2003)

8. Eide, A.W., Pickering, J.B., Yasseri, T., Bravos, G., Følstad, A., Engen, V., Walland, P., Meyer, E.T., Tsvetkova, M.: Human-Machine Networks: Towards a Typology and Profiling Framework. In: HCI International. Springer, Toronto, Canada (2016)

9. Leonardi, P.M.: Materiality, Sociomateriality, and Socio-Technical Systems: What Do These Terms Mean? How Are They Related? Do We Need Them? In: Leonardi, P.M., Nardi, B.A., and Kallinikos, J. (eds.) Materiality and Organizing: Social Interaction in a Technological World. pp. 25–48. Oxford University Press, Oxford (2012)

10. Smart, P.R., Simperl, E., Shadbolt, N.: A Taxonomic Framework for Social Machines, http://eprints.soton.ac.uk/362359/1/SOCIAM Classificationv2.pdf, (2014)

11. Tsvetkova, M., Yasseri, T., Meyer, E.T., Pickering, J.B., Engen, V., Walland, P., Lüders, M., Følstad, A., Bravos, G.: Understanding Human-Machine Networks: A Cross-Disciplinary Survey. arXiv Prepr. (2015)

12. Bandura, A.: Social Cognitive Theory: An Agentic Perspective. Annu. Rev. Psychol. 52, 1–26 (2001)

13. Giddens, A.: The Constitution of Society: Outline of the Theory of Structuration., (1984)

14. Friedman, B., Kahn, P.H.: Human Agency and Responsible Computing: Implications for Computer System Design. J. Syst. Softw. 17, 7–14 (1992)

15. Rose, J., Truex, D.: Machine Agency as Perceived Autonomy: An Action Perspective. In: Organizational and Social Perspectives on Information Technology. pp. 371–388. Springer (2000)

16. Rose, J., Jones, M., Truex, D.: The Problem of Agency: How Humans Act, How Machines Act. In: International workshop on Action in Language, Organisations and Information Systems (ALOIS-2003) (2003)

17. Nass, C.I., Lombard, M., Henriksen, L., Steuer, J.: Anthropocentrism and computers. Behav. Inf. Technol. 14, 229–238 (1995)

18. Bandura, A.: Self-Efficacy: The Exercise of Control. Freeman, New York (1997)

19. Boshmaf, Y., Muslukhov, I., Beznosov, K., Ripeanu, M.: The Socialbot Network: When Bots Socialize for Fame and Money. When Bots Social. Fame Money. 93 (2011)

20. Chu, Z., Gianvecchio, S., Wang, H., Jajodia, S.: Who is Tweeting on Twitter: Human, Bot, or Cyborg? Proc. 26th Annu. Comput. Secur. Appl. Conf. - ACSAC '10. 21 (2010)

21. von Ahn, L., Maurer, B., McMillen, C., Abraham, D., Blum, M.: reCAPTCHA: human-based character recognition via Web security measures. Science. 321, 1465–8 (2008)

22. Picard, R.W.: Affective Computing. (1995)

23. Mayer, R.C., Davis, J.H., Schoorman, F.D.: An Integrative Model of Organizational Trust. Acad. Manag. Rev. 20, 709–734 (1995)

24. Schoorman, F.D., Mayer, R.C., Davis, J.H.: An Integrative Model of Organizational Trust: Past, Present, and Future. Acad. Manag. Rev. 32, 344–354 (2007)

25. Nass, C., Moon, Y.: Machines and Mindlessness: Social Responses to Computers. J. Soc. Issues. 56, 81–103 (2000)

26. King, W.R., He, J.: A meta-analysis of the technology acceptance model. Inf. Manag. 43, 740–755 (2006)

27. Corritore, C.L., Kracher, B., Wiedenbeck, S.: On-line trust: concepts, evolving themes, a model. Int. J. Hum. Comput. Stud. 58, 737–758 (2003)

28. Wiegmann, D. a., Rich, A., Zhang, H.: Automated diagnostic aids: The effects of aid reliability on users' trust and reliance. Theor. Issues Ergon. Sci. 2, 352–367 (2001)

29. Dutton, W.H., Shepherd, A.: Trust in the Internet as an experience technology, (2006)

30. Scraton, P.: Policing with Contempt: The Degrading of Truth and Denial of Justice in the Aftermath of the Hillsborough Disaster. J. Law Soc. 26, 273–297 (1999)

31. Taylor, S.P.: The Hillsborough stadium disaster: 15 Apirl 1989: inquiry by the Rt hon Lord justic Taylor: final report: present to Parliament by the Secretary of State for the Home department by command of Her Majesty January 1990. (1990)

32. Drury, J., Reicher, S.: Collective action and psychological change: The emergence of new social identities. Br. J. Soc. Psychol. 39, 579–604 (2000)

33. Nicholson, C.E., Roebuck, B.: The investigation of the Hillsborough disaster by the Health and Safety Executive. Saf. Sci. 18, 249–259 (1995)

34. Donald, I., Canter, D.: Intentionality and fatality during the King's Cross underground fire. Eur. J. Soc. Psychol. 22, 203–218 (1992)

35. Canter, D. V: Fires and human behaviour. John Wiley & Sons (1980)

36. Drury, J., Cocking, C.: The mass psychology of disasters and emergency evacuations: A research report and implications for practice. (2007)

37. Pickering, J.B.: TRust in IT: Factors, metRics, Models, http://www.itutility.ac.uk/2014/10/30/trust-in-it-factors-metrics-models/

38. Information Commissioner's Office: Key Definitions of the Data Protection Act. (2015)

39. European Commission: DIRECTIVE 95/46/EC OF THE EUROPEAN PARLIAMENT AND OF THE COUNCIL. (1995)

40. Poyan Sandnell: Dreaming of a connected (and smart) home, http://www.ericsson.com/thinkingahead/the-networked-society-blog/2014/12/22/dreaming-connected-smart-home/

41. Howe, J.: The rise of crowdsourcing. Wired Mag. (2006)

42. Surowiecki, J.: The Wisdom of Crowds. Anchor (2005)